\documentclass[letterpaper,notoc]{CECS}
\title{Supersymmetry of gravitational ground states}
\author{Rodrigo  Aros$^1$,  Cristi\'an Mart\'{\i}nez, Ricardo Troncoso and Jorge Zanelli
\footnote{{\it E-mail:} {\tt  rod, martinez, ratron, jz@cecs.cl}}
\\ Centro de Estudios Cient\'{\i}ficos (CECS), Casilla 1469, Valdivia, Chile
\\$^1$Universidad Nacional Andr\'{e}s Bello, Sazie 2320, Santiago,
Chile }

\preprint{{\tiny CECS-PHY-02/02}}

\abstract{ A class of black objects which are solutions of pure gravity with negative cosmological
constant are classified through the mapping between the Killing spinors of the ground state and
those of the transverse section. It is shown that these geometries must have transverse sections
of constant curvature for spacetime dimensions $d$ below seven. For $d\geq 7$, the transverse
sections can also be Euclidean Einstein manifolds. In even dimensions, spacetimes with transverse
section of nonconstant curvature exist only in $d=8$ and $10$. This classification goes beyond
standard supergravity and the eleven dimensional case is analyzed. It is shown that if the
transverse section has negative scalar curvature, only extended objects can have a supersymmetric
ground state. In that case, some solutions are explicitly found whose ground state resembles a
wormhole.}

\begin{document}
\section{Introduction}

The presence of a negative cosmological constant allows the existence of black holes with
topologically non-trivial transverse sections \cite{Lemos:1995xp, Vanzo:1997gw,Brill:1997mf}. The
features of these geometries have been extensively studied, see e.g., \cite{Birmingham:1998nr,
Cai:1998vy,Caldarelli:1998hg, Emparan:1999pm, Galloway:1999bp, Aros:2000ij}. The simplest solution
of the Einstein equations with negative cosmological constant of this\ kind in $d\geq 4$
dimensions, is described by the line element\footnote{The cosmological constant is given by
$\Lambda =-l^{-2}(d-1)(d-2)/2$.}
\begin{equation}\label{Top-BH-Einstein}
ds^{2} =-\left( \gamma +\frac{r^{2}}{l^{2}}-\frac{\mu }{r^{d-3}}\right) dt^{2}+
\frac{dr^{2}}{\left( \gamma +\frac{r^{2}}{l^{2}}-\frac{\mu }{r^{d-3}}\right)} +r^{2}d\sigma
_{\gamma }^{2},
\end{equation}
where the constant $\mu $ is proportional to the mass. Here $d\sigma
_{\gamma }^{2}=\hat{g}_{ij}(y)dy^{i}dy^{j}$ is the line element of the $%
(d-2) $-dimensional transverse section $\Sigma _{\gamma }$, which is a Einstein manifold of
Euclidean signature,
\begin{equation}
\hat{R}_{ij}=\gamma (d-3)\hat{g}_{ij}.  \label{EinsteinTransverse}
\end{equation}
The constant $\gamma $ has been normalized to $\pm 1,0$ by a suitable coordinate rescaling.

The electrically and magnetically charged extensions of (\ref {Top-BH-Einstein}) are also known
\cite{Brill:1997mf}. The existence of an event horizon is ensured if $\Sigma _{\gamma }$ is a
compact and orientable surface. The Schwarzschild-AdS geometry is recovered when the transverse
section is the unit sphere ($\gamma =1$).

Note that the configurations (\ref{Top-BH-Einstein}) are asymptotically locally AdS spacetimes
only if the transverse section $\Sigma _{\gamma }$ has constant curvature, namely, the curvature
two-form\footnote{This condition is automatically satisfied in four and five dimensions.}
satisfies $\hat{R}^{mn}=\gamma \hat{e}^{m}\mbox{\tiny $\wedge$ }\hat{e}^{n}$. In particular, this
means that $\Sigma _{\gamma }$ is locally isomorphic to the sphere $S^{n}$, the hyperbolic
manifold $H^{n}$, or the Euclidean space $\mathbb{R}^{n}$. Furthermore, the Killing-Hopf theorem
states that any $n$-dimensional complete connected Riemannian manifold of Euclidean signature and
constant curvature $\gamma $ (for $n\geq 2$), has one of the following forms\ (see, e. g.
\cite{Wolf1984})
\[
\begin{tabular}{lll}
$S^{n}/\Gamma $ & with $\Gamma \subset O(n+1)$ & if $\gamma >0$ \\
$H^{n}/\Gamma $ & with $\Gamma \subset O(n,1)$ & if $\gamma <0$ \\
$\mathbb{R}^{n}/\Gamma $ & with $\Gamma \subset ISO(n)$ & if $\gamma =0$%
\end{tabular}
\]
where $\Gamma $ is a freely acting discrete subgroup (i.e., without fixed points).

Asymptotically locally anti de Sitter black holes are nontrivial examples for testing the AdS/CFT
correspondence \cite{Aharony:1999ti}. Their asymptotic regions provide inequivalent background
spacetimes where the corresponding dual thermal CFT is realized \cite
{Witten:1998zw,Birmingham:1998nr,Emparan:1999pm}. Thus, CFT's defined on $%
S^{1}\times (\mathbb{R}^{d-2}/\Gamma )$, $S^{1}\times (S^{d-2}/\Gamma )$ or $%
S^{1}\times (H^{d-2}/\Gamma )$ are connected with black holes in the bulk for $\gamma =0$, $1$, or
$-1$, respectively. Black holes with topologically non-trivial transverse sections exist also for
gravitation theories containing higher powers of the curvature. Their relationship with thermal
CFT has also been explored \cite{Aros:2000ij}.

Since we assume a nonvanishing cosmological constant, and the geometry of $%
\Sigma _{\gamma }\,$is not necessarily spherical, these solutions do not satisfy the hypotheses of
the standard positivity energy theorems for gravity and supergravity
\cite{Deser:1977hu,Schoen:1979rg,Witten:1981mf}. As a step towards a proof of an energy bound, we
analyze the existence of supersymmetric ground states. This requirement not only ensures the
stability of the ground state, but also impose severe restrictions on the geometry of $\Sigma
_{\gamma }$, allowing a classification of the ground states.

The classification of Euclidean manifolds admitting Killing spinors is obtained from their
transformation properties under parallel transport along a closed loop, that is, from the holonomy
group of the manifold. The possible holonomy groups, in turn, were classified by Berger
\cite{Berger}.

In the following, the conditions for the family of spacetimes of the form (\ref{Top-BH-Einstein})
to admit Killing spinors is addressed. In Section \ref{KillingBegin}, the Killing spinors of
spacetime are explicitly obtained, and they are completely determined by the Killing spinors of
the transverse section. This occurs for $\mu =0$\ only.

In Section \ref{Classification}, the supersymmetric ground states are analyzed for the different
values of $\gamma $. Since a manifold of positive scalar curvature ($\gamma =1$) is necessarily
compact, the asymptotic region of non-rotating localized distributions of matter with a
supersymmetric ground state are classified.

For $\gamma =-1$ supersymmetry requires the transverse section to be noncompact. All
supersymmetric geometries of the form (\ref{Top-BH-Einstein}) with noncompact transverse section
of negative scalar curvature are classified. The transverse sections of these geometries contain a
Ricci flat submanifold which determines the number of supersymmetries of spacetime. Warped black
brane solutions are found whose supersymmetric ground state resembles a wormhole.

For $\gamma =0$, the study is restricted to a particular class of Ricci flat transverse sections
allowing the existence of proper black objects, which are classified demanding the existence a
supersymmetric ground state.

In Section \ref{Elevendimensions} it is shown that this classification of supersymmetric ground
states in standard supergravity also applies to eleven dimensional AdS Supergravity
\cite{Troncoso:1998va,Troncoso:1997me}.

\section{Killing spinors and the transverse section}

\label{KillingBegin}

The spacetimes (\ref{Top-BH-Einstein}) can be viewed as solutions of standard supergravity
theories with negative cosmological constant (see \emph{e. g.}\cite{Salam:1989fm}). These
configurations are left invariant under supersymmetry transformations, $\delta \psi =\nabla
\epsilon $, provided $\epsilon $ is a global solution of the Killing spinor equation
\begin{equation}
\nabla \epsilon :=\left( d+A\right) \epsilon =0,  \label{KillingAdS}
\end{equation}
where
\begin{equation}
A=\frac{1}{4}\omega ^{ab}\Gamma _{ab}+\frac{1}{2l}e^{a}\Gamma _{a}\;, \label{A}
\end{equation}
and $l$ is the AdS radius. The one-form $A$ can be regarded as a connection for the AdS group
$SO(d-1,2)$, whose generators are expressed in terms of
Dirac matrices as $J_{ab}=\frac{1}{2}\Gamma _{ab}$ and $J_{a}=\frac{1}{2}%
\Gamma _{a}$, and whose curvature $F=dA+A\mbox{\tiny $\wedge$}A$ is \footnote{%
Here $\omega ^{ab}$ is the Lorentz connection one-form, $e^{a}$ is the vielbein, $R^{ab}=d\omega
^{ab}+\omega _{\,\,c}^{a}\mbox{\tiny $\wedge$}\omega ^{cb}$ is the curvature two-form and
$T^{a}=de^{a}+\omega ^{a}_{\,\,b} \mbox{\tiny $\wedge$}e^{b}$ is the torsion.}
\begin{equation}
F=\frac{1}{2}(R^{ab}+\frac{1}{l^{2}}e^{a}\mbox{\tiny $\wedge$ } e^{b})J_{ab}+\frac{1}{l}%
T^{a}J_{a}.  \label{F}
\end{equation}

The integrability condition for equation (\ref{KillingAdS}) reads
\begin{equation}
\nabla \nabla \epsilon =F\epsilon =0.  \label{Integrability}
\end{equation}
As the torsion vanishes, the curvature for the metric (\ref{Top-BH-Einstein}%
) is
\[
R^{ab}+\frac{1}{l^{2}}e^{a}\mbox{\tiny $\wedge$ } e^{b}=\left\{
\begin{tabular}{l}
$\mu \frac{(d-2)(d-3)}{2r^{d-1}}e^{0}\mbox{\tiny $\wedge$ } e^{1}$ \\
\\
$-\mu \frac{(d-3)}{2r^{d-1}}e^{0}\mbox{\tiny $\wedge$ } e^{m}$ \\
\\
$-\mu \frac{(d-3)}{2r^{d-1}}e^{1}\mbox{\tiny $\wedge$ } e^{m}$ \\
\\
$\hat{R}^{mn}-\gamma \hat{e}^{m}\mbox{\tiny $\wedge$ } \hat{e}^{n}+\mu \frac{1}{r^{d-1}}%
e^{m}\mbox{\tiny $\wedge$ } e^{n}$%
\end{tabular}
\right.
\]
The integrability condition (\ref{Integrability}) is satisfied only if $\mu =0$ and
\[
(\hat{R}^{mn}-\gamma \hat{e}^{m}\mbox{\tiny $\wedge$ } \hat{e}^{n})\Gamma _{mn}\epsilon =0\;.
\]

Thus, the existence of Killing spinors shall be investigated for massless geometries whose line
element is of the form

\begin{equation}
ds^{2}=-\left( \gamma +r^{2}/l^{2}\right) dt^{2}+\frac{dr^{2}}{\left( \gamma +r^{2}/l^{2}\right)
}+r^{2}d\sigma _{\gamma }^{2}\;.  \label{Top-BH-vacuum}
\end{equation}

Choosing the frame as

\[
\begin{array}{lll}
e^{0}=f(r)dt, & e^{1}=f(r)^{-1}dr, & e^{m}=r\hat{e}^{m},
\end{array}
\]
where $f(r)=\sqrt{r^{2}/l^{2}+\gamma }$, and $\hat{e}^{m}(y)$ is the vielbein of the transverse
section $\Sigma _{\gamma }$, the connection one-form $A$ reads
\[A =\left( \frac{1}{2l}f(r)-\frac{r}{2l^{2}}\Gamma _{1}\right) \Gamma
_{0}dt+\frac{dr}{2lf(r)}\Gamma _{1} +\frac{1}{4}\hat{\omega}^{mn}\Gamma _{mn}+\left(
\frac{r}{l}-f(r)\Gamma _{1}\right) \frac{1}{2}\hat{e}^{m}\Gamma _{m},
\]
where $\hat{\omega}^{mn}(y)$ is the spin connection of the transverse section. In this frame, the
solution of the Killing spinor equation (\ref {KillingAdS}) reads
\begin{equation}
\epsilon =e^{-\frac{\Gamma _{1}}{2}\ln \left( r/l+\sqrt{\gamma +r^{2}/l^{2}}%
\right) }e^{-\frac{t}{l}P_{0}}\eta \;,  \label{Solution}
\end{equation}
where $\eta (y)$ satisfies
\begin{equation}
\left( d+\hat{A}_{\gamma }\right) \eta =0\;,  \label{KillingTransverse}
\end{equation}
and
\begin{equation}
\hat{A}_{\gamma }=\frac{1}{2}\hat{\omega}^{mn}J_{mn}+\hat{e}^{m}P_{m}. \label{hatA}
\end{equation}
Here
\begin{eqnarray}
P_{m} &=&\frac{1}{2}(P_{-}-\gamma P_{+})\Gamma _{m},  \nonumber \\
P_{0} &=&\frac{1}{2}\left( P_{-}+\gamma P_{+}\right) \Gamma _{0},  \label{Ps}
\end{eqnarray}
with $P_{\pm }:=\frac{1}{2}(1\pm \Gamma _{1})$.

Note that since $[P_{m},P_{n}]=-\gamma J_{mn}$, the set $\left\{P_{n},J_{mn}
\right\} $ forms a \emph{reducible} representation for $SO(d-1)$, $SO(d-2,1)$%
, or $ISO(d-2)$ depending on whether $\gamma =1,-1,$ or $0$, respectively. It is simple to express
$\eta $ in terms of irreducible $(d-2)$ -dimensional spinors for each case (see below).

In this way, the problem of finding a Killing spinor $\epsilon $ for the spacetime
Eq.(\ref{Top-BH-vacuum}), has been reduced to that of finding a globally defined spinor $\eta $ on
the transverse section. Thus, one can formulate the following lemma\medskip

\textbf{Lemma:}\label{Lemma}\textit{\ Killing spinors for the geometries (%
\ref{Top-BH-Einstein}) exist only for }$\mathit{\mu =0}$\textit{, and are completely determined by
the Killing spinors of the transverse section.\medskip }

This situation is analogous to the case of conifold geometries, where the Killing spinors of an
Einstein manifold $X$ of positive scalar curvature are related to the Killing spinors of the cone
over $X$, which is a Ricci flat manifold (see \cite{Figueroa-O'Farrill:1999va}). Here, the Killing
spinors of the transverse section $\Sigma _{\gamma }$, which is an Einstein manifold of scalar
curvature $\hat{R}=\gamma (d-2)(d-3)$, determine the Killing spinors of the spacetime
(\ref{Top-BH-vacuum}), which is an Einstein manifold of negative scalar curvature.

Since complete, connected, irreducible Riemannian manifolds of Euclidean signature admitting
Killing spinors, have been classified \cite {Wang1989,Bar93,Baum731989} the above lemma allows the
classification of the black objects (\ref{Top-BH-Einstein}) with a supersymmetric ground state.
This is discussed in the following section.

\section{Classification of ground states}

\label{Classification}

Since spinors may transform nontrivially under parallel transport along a closed loop, the maximal
number of supersymmetries of a Euclidean manifold is determined by its holonomy group
\cite{Wang1989}, which are classified by Berger's theorem.

Killing spinors of a simply connected, complete and irreducible Einstein manifold $X$ of positive
scalar curvature were classified in \cite{Bar93} by using the conifold mapping between $X$ and the
cone over $X$ which is Ricci flat. Analogously, there is a one-to-one correspondence between
Killing
spinors of an Einstein manifold of negative scalar curvature, $\Sigma _{-1}$%
, with the supersymmetries of Ricci flat manifold one dimension below \cite {Baum731989}.

Smooth manifolds can be obtained by making quotients of a symmetric space by discrete subgroups
without fixed points. These quotients, in general, make the geometry non simply connected and can
introduce noncontractible loops which might further reduce the number of supersymmetries.

If the transverse section is a Ricci flat reducible space, then Eq.(%
\ref{KillingTransverse}) decomposes into the Killing spinor equation for each of its
irreducible factors. Hence, $\Sigma _{0}$ admits global Killing spinors if and only if each factor does.

In what follows, the three cases $\gamma =\pm 1,0$, are analyzed separately.

\subsection{Positive curvature transverse section}

\label{PositiveCurvature}

In order to express $\eta $ in terms of its irreducible parts, it is useful to introduce the
projectors $\mathcal{Q}_{\pm }=\left( \frac{1}{2}\pm iP_{0}\right) $ which commute with the
one-form $\hat{A}$ defined by Eq.(\ref {hatA}). This allows splitting Eq.(\ref{KillingTransverse})
as
\begin{equation}
d\eta _{\pm }+\hat{A}_{\pm }\eta _{\pm }=0\;,  \label{SplitEtaSpherical}
\end{equation}
where $\hat{A}_{\pm }=\mathcal{Q}_{\pm }\hat{A}$ are irreducible representations of $\hat{A}$,
acting on $\eta _{\pm }=\mathcal{Q}_{\pm }\eta $, which are genuine spinors of the transverse
section. When the dimension of $\Sigma _{1}$ is odd, then both representations are inequivalent,
unlike
the even dimensional case\footnote{%
\medskip When $\Sigma _{1}\,$is $2n$-dimensional, then $-\gamma ^{i}=\gamma
_{2n+1}\gamma ^{i}\gamma _{2n+1}$, with $\gamma _{2n+1}=(i)^{n}\gamma _{1}\gamma _{2}...\gamma
_{2n}$.}. Choosing the following representation for
the Dirac matrices \footnote{%
Here and henceforth, $\sigma _{i}$ are the Pauli matrices and $\gamma _{m}$ satisfy
$(d-2)$-dimensional Clifford algebra, $\{\gamma _{m},\gamma _{n}\}=2\delta _{mn}$, with
$m=2,...,d-1$.}: $\Gamma _{0}=i\sigma
_{z}\otimes \mathbb{I}$, $\Gamma _{1}=\sigma _{x}\otimes \mathbb{I}$, $%
\Gamma _{m}=\mathbb{\sigma }_{y}\otimes \gamma _{m}$, Eq.(\ref {SplitEtaSpherical}) reads
\begin{equation}
d\eta _{\pm }+\left( \pm \frac{i}{2}\gamma _{m}\hat{e}^{m}+\frac{1}{4}\hat{%
\omega}^{mn}\gamma _{mn}\right) \eta _{\pm }=0\;, \label{SplitEtaRepSpherical}
\end{equation}
which means that $\eta _{\pm }$ must be a globally defined Killing spinor on $\Sigma _{1}$. Let
$N_{\pm }$ be the maximum possible number of solutions of type $\eta _{\pm }$. Then, using the
classification of positive scalar curvature Euclidean manifolds admitting Killing spinors in
\cite{Bar93}, one can formulate the following theorem:\medskip

\textbf{Theorem 1.} Let $\mathcal{M}$\ be a $d$-dimensional manifold of the
form (\ref{Top-BH-Einstein}) with $\gamma =1$, whose transverse section $%
\Sigma _{1}$ is a simply connected, complete and irreducible Riemannian manifold of positive
scalar curvature. If $\mathcal{M}$ possesses a supersymmetric ground state, then $\mathcal{M}$\
can be either

\textbf{(i)} a $d$-dimensional Schwarzschild-AdS spacetime, whose ground
state is AdS, admitting the maximum number of supersymmetries, namely $%
2^{[d/2]}$,

\textbf{(ii)} an $8$-dimensional black hole whose transverse section is a nearly K\"{a}hler
manifold, and its ground state admits one Killing spinor of each type ($N_{+}=N_{-}=1$), or

\textbf{(iii)} an odd-dimensional black hole with $d\geq 7$, whose transverse section geometry and
the corresponding maximum number of Killing spinors of its ground state are given by the following
table,
\begin{center}
\textbf{Table 1}
\medskip

\begin{tabular}{|c|c|c|}
\hline $d$ & $\Sigma _{1}$ & $(N_{+},N_{-})$ \\ \hline\hline  $7$ & Sasaki-Einstein & $(1,1)$ \\
\hline$9$ &
\begin{tabular}{c}
3-Sasaki \\
Sasaki-Einstein \\
Nearly
parallel $G_{2}$
\end{tabular}
&
\begin{tabular}{c}
$\left( 3,0\right) $ \\
$(2,0)$ \\
$(1,0)$%
\end{tabular}
\\ \hline
$11$ & Sasaki-Einstein & $(1,1)$ \\ \hline
\end{tabular}
\end{center}
\medskip

Since, under the assumptions of the above theorem, the transverse section $%
\Sigma _{1}$ is necessarily a compact manifold, this theorem classifies the asymptotic region of
localized nonrotating distributions of matter with a supersymmetric ground state.

From a mathematical point of view, the Killing spinor equation can be solved regardless the
existence of a supergravity theory. Indeed, the transverse section for $d=4k-1>11$ dimensions, can
be a Sasaki-Einstein manifold admitting $(1,1)$ Killing spinors; and for $d=4k+1>9$ the surface
$\Sigma
_{1}$ can be either a Sasaki-Einstein or a 3-Sasaki manifold admitting $%
(2,0) $ and $\left( k+1,0\right) $ Killing spinors respectively.

As mentioned above, performing identifications on these transverse sections, break some
supersymmetries in general. For instance, consider the smooth
quotients of the sphere which have been fully classified (see \emph{e. g.}%
\cite{Wolf1984}). In even dimensions, real projective spaces, --that is, the
sphere with antipodal points identified, $\mathbb{RP}^{2n}=S^{2n}/\mathbb{Z}%
_{2}$--, are the only possible smooth quotients. However, these are not orientable manifolds and
therefore they cannot correspond to the event horizon of a black hole.

In odd dimensions, smooth quotients of the form $\Sigma _{1}=S^{2n-1}/\Gamma $ are always
orientable, and among them, there are some interesting cases with unbroken supersymmetries (see
\emph{e. g. }Ref. \cite{Kachru:1998ys}). These correspond to the transverse section of topological
black holes whose ground states are locally AdS spaces with unbroken supersymmetries. For
instance, the real projective space $\mathbb{RP}^{2n-1}$ admits $(2^{n-1},0)$ Killing spinors,
provided $n$ is even \cite{Bar93}. Thus, in nine dimensions, if $\Sigma _{1}=\mathbb{RP}^{7}$, the
spacetime admits 8 Killing
spinors which are eigenstates of $\mathcal{Q}_{+}$ and zero eigenstates of $%
\mathcal{Q}_{-}$ (or vice-versa).

\subsection{Negative curvature transverse section}

\label{NegativeCurvature}

The line element (\ref{Top-BH-Einstein}) can be viewed as the exterior geometry of a localized
non-rotating distribution of matter, provided the transverse section $\Sigma _{\gamma }$ is a
compact Euclidean Einstein manifold (see Eq. (\ref{EinsteinTransverse})). However, as can be
easily seen, a compact transverse section $\Sigma _{\gamma }$ admits no Killing spinors for
$\gamma =-1$. Consider vector $\xi ^{m}$ which satisfies the Killing equation on $\Sigma _{-1}$.
Then, the following identity holds
\begin{equation}
\int_{\Sigma _{-1}}\nabla _{m}\xi _{n}\nabla ^{m}\xi ^{n}\sqrt{\hat{g}}%
d^{d-2}x=\gamma (d-3)\int_{\Sigma _{-1}}\xi _{m}\xi ^{m}\sqrt{\hat{g}}%
d^{d-2}x\;.  \label{Identity}
\end{equation}
Since $\Sigma _{\gamma }$ has Euclidean signature, the left hand side of Eq. (\ref{Identity}) is
nonnegative, and therefore, for $\gamma =-1$, $\xi ^{m}$ necessarily vanishes. Furthermore, since
for any Killing spinor $\eta $, the vector field $\xi ^{m}:=\bar{\eta}\gamma ^{m}\eta $ would be
an isometry, one concludes that any compact Euclidean Einstein manifold with negative scalar
curvature cannot have either Killing vectors, or Killing spinors.

As a consequence of this, and by virtue of the lemma in section \ref{Lemma}, black holes with
compact transverse sections cannot be BPS states regardless the value of the mass. In particular,
in this case, the solution with $\mu
=0 $ describes a black hole with horizon\ radius $r_{+}=l$ and temperature $%
\beta =2\pi l$. Now, since the specific heat is positive, this state could
decay by Hawking radiation into black holes with $0>\mu \geq \mu _{c}=-\frac{%
l^{d-3}}{G}\sqrt{\frac{(d-3)^{d-3}}{(d-1)^{d-1}}}$ \cite{Aros:2000ij}.

Therefore, demanding that the metric (\ref{Top-BH-vacuum}) admit globally defined Killing spinors,
leads\ one to consider geometries with noncompact transverse sections $\Sigma _{-1}$. Noncompact
Riemannian manifolds, which are complete, connected and irreducible, of Euclidean signature and
negative scalar curvature, admitting Killing spinors have been classified \cite {Baum731989}.\
Since the line element is isometric to
\begin{equation}
d\sigma _{-1}^{2}=\frac{1}{z^{2}}(dz^{2}+h_{ij}dx^{i}dx^{j})\;, \label{Sigma-1}
\end{equation}
where $h_{ij}$ is the metric of a $d-3$-dimensional complete connected Ricci-flat manifold
admitting Killing spinors, the classification of these spaces is obtained from the classification
of Ricci flat manifolds in Ref. \cite{Wang1989}.

In complete analogy with case $\gamma =1$, the metrics of the form (\ref {Top-BH-vacuum}) with
$\gamma =-1$ admit Killing spinors given by Eq.(\ref {Solution}). The irreducible components of
$\hat{A}$\ and $\eta $ are also given by $\hat{A}_{\pm }=\mathcal{Q}_{\pm }\hat{A}$\ and $\eta
_{\pm }=
\mathcal{Q}_{\pm }\eta $, respectively\footnote{%
When the dimension of $\Sigma _{-1}$\ is even both representations are equivalent, unlike the odd
case.}, where now $\mathcal{Q}_{\pm }:=\left( \frac{1}{2}\pm P_{0}\right) $.

Using the representation for the Dirac matrices in which $\Gamma
_{0}=i\sigma _{y}\otimes \mathbb{I}$, $\Gamma _{1}=\sigma _{x}\otimes %
\mathbb{I}$, $\Gamma _{m}=\sigma _{z}\otimes \gamma _{m}$ for $\gamma =-1$, one finds that $\eta
_{\pm }$ satisfy,
\begin{equation}
d\eta _{\pm }+\left( \pm \frac{1}{2}\gamma _{m}\hat{e}^{m}+\frac{1}{4}\hat{%
\omega}^{mn}\gamma _{mn}\right) \eta _{\pm }=0\;.  \label{SplitEtaRep}
\end{equation}

Let $N_{\pm }$ be the maximum possible number of solutions of type $\eta _{\pm }$. Then, since
negative scalar curvature Euclidean manifolds admitting Killing spinors are classified in
\cite{Baum731989}, the following theorem holds:\medskip

\textbf{Theorem 2.} Let $\mathcal{M}_{0}$\ be a $d$-dimensional manifold of the form
(\ref{Top-BH-vacuum}) with $\gamma =-1$, admitting Killing spinors, whose transverse section
$\Sigma _{-1}$ is a noncompact, connected, complete and irreducible manifold of negative scalar
curvature. Let $\Xi $ be the complete, connected Ricci-flat submanifold described by $h_{ij}$ in
Eq.(\ref {Sigma-1}). If $\mathcal{M}_{0}$ possesses a supersymmetric ground state, then
$\mathcal{M}_{0}$\ can be either

\textbf{(i)} a $d$-dimensional portion of AdS whose transverse section is $%
H^{d-2}$, which admits $2^{[(d-2)/2]}$ Killing spinors,

\textbf{(ii)} a $10$-dimensional manifold where $\Xi $ is a $7$-dimensional space with $G_{2}$
holonomy admitting only one Killing spinor, or

\textbf{(iii)} an odd-dimensional manifold with $d\geq 7$, where the geometry, holonomy and
corresponding maximal number of Killing spinors of $\Xi$ is given by the following table,
\begin{center}
\textbf{Table 2}
\medskip

\begin{tabular}{|c|c|c|c|}
\hline $d$ & $\Xi $ & Hol($\Xi$) & $(N_{+},N_{-})$ \\ \hline\hline $7$ &
\begin{tabular}{c}
hyperk\"{a}hler \\
Calabi-Yau
\end{tabular}
&
\begin{tabular}{c}
$Sp(2)$ \\
$SU(2)$%
\end{tabular}
&
\begin{tabular}{c}
$(2,0)$ \\
$(2,0)$%
\end{tabular}
\\ \hline
$9$ & Calabi-Yau & $SU(3)$ & $(1,1)$ \\ \hline $11$ &
\begin{tabular}{c}
hyperk\"{a}hler \\
Calabi-Yau \\
Parallel Spin$_{7}$%
\end{tabular}
&
\begin{tabular}{c}
$Sp(4)$ \\
$SU(4)$ \\
$Spin(7)$%
\end{tabular}
&
\begin{tabular}{l}
$(3,0)$ \\
$(2,0)$ \\
$(1,0)$%
\end{tabular}
\\ \hline
\end{tabular}
\end{center}
\medskip

As in the previous case, independently from the existence of supergravity,
for $d=4k+1>9$, the surface $\Xi $ can be a Calabi-Yau manifold with $%
SU(2k-1)$ holonomy which admits $(1,1)$ Killing spinors,\ while for $%
d=4k-1>11$, $\Xi $\ can be either a Calabi-Yau manifold with $SU(2k-2)$
holonomy admitting $(2,0)$ Killing spinors, or hyperk\"{a}hler with $%
Sp(2k-2) $ holonomy which admits $(k,0)$ Killing spinors.

\subsubsection{Supersymmetry and extended objects}

Remarkably, for $\gamma =-1$, the requirement of supersymmetry implies\ noncompactness of the
transverse section $\Sigma _{-1}$. This in turn means that only extended objects can have a
supersymmetric ground state.

Let us consider for example, the four dimensional case, whose transverse section $\Sigma _{-1}$ is
a two-dimensional surface of negative constant curvature. If $\Sigma _{-1}=H_{2}$, the metric
(\ref{Top-BH-vacuum}) describes a portion of AdS$_{4}$ instead of a topological black hole \cite
{Vanzo:1997gw,Brill:1997mf}. However, if one considers\ a quotient of the
form $\Sigma _{-1}=H_{2}/\Gamma $, which is topologically a cylinder ($%
\mathbb{R}\times S^{1}$), then the metric (\ref{Top-BH-Einstein}) with $%
\gamma =-1$ describes a warped black string which may, or may not, possess a supersymmetric ground
state, depending on $\Gamma $, as it is seen in the following examples
\vskip 15mm

$\circ $ \textit{Non supersymmetric  ``ground state''}\medskip

Consider $\Sigma _{-1}$ described by the metric,
\begin{equation}
d\sigma _{-1}^{2}=d\zeta^{2}+\cosh ^{2} \zeta d\varphi ^{2}\;, \label{nonsusy-cyllinder}
\end{equation}
with $-\infty <\zeta<\infty $ and $0<\varphi \leq \alpha $, obtained by an
identification on $H_{2}$ along the boost $\Gamma =\alpha \partial _{\phi }$%
. The solution of the Killing spinor equation (\ref{SplitEtaRep}) is given by
\begin{equation}
\eta =\exp \left( -\frac{\zeta}{2}\sigma _{2}\right) \exp \left( -\frac{\varphi }{2}\sigma
_{3}\right) \eta _{0}\;.
\end{equation}
However, $\eta $ is not globally defined because $\eta (\zeta,\alpha )\neq \pm \eta (\zeta,0)$,
and therefore this space admits no Killing spinors. For $\mu =0$ the spacetime is locally AdS and
has an event horizon at $r_{+}=l$. This solution is not supersymmetric and could decay by Hawking
radiation to a state with $\mu <0$. Cosmic censorship requires the ground state to be the solution
with $\mu _{c}=-\frac{l}{3\sqrt{3}G}$, which is not supersymmetric either.\medskip

$\circ $ \textit{Supersymmetric ground state}\medskip

Consider $\Sigma _{-1}$ to be the Poincar\'{e} upper half\textit{\ cylinder}
\begin{equation}
d\sigma _{-1}^{2}=\frac{1}{z^{2}}\left( dz^{2}+d\varphi ^{2}\right) , \label{WarpedH2}
\end{equation}
with $0<z<\infty $ and $0<\varphi \leq \alpha $, which is obtained by wrapping $H_{2}$ along the
isometry $\Gamma =\alpha \partial _{\varphi }$. Then the solution of the Killing equation is
\begin{equation}
\eta (z,\varphi )=\exp \left( -\frac{\ln (z)}{2}\sigma _{2}\right) \left( I- \frac{\varphi
}{2}\sigma _{3}(I+\sigma _{2})\right) \eta _{0}. \label{WarpedEta}
\end{equation}
This is a globally defined spinor provided $(I+\sigma _{2})\eta _{0}=0$. Thus, in four dimensions,
the line element
\begin{equation}\label{Warped Black String}
ds^{2} =-\left( r^{2}/l^{2}-\frac{\mu }{r}-1\right) dt^{2}+ \frac{dr^{2}}{\left(
r^{2}/l^{2}-\frac{\mu }{r}-1\right) }+\frac{r^{2}}{z^{2}}\left( dz^{2}+d\varphi ^{2}\right),
\end{equation}
describes a warped black string of mass $M=V_{2}\frac{\mu }{8\pi }$, where $V_{2}$ is the area of
the transverse section.

For $\mu =0$ the metric (\ref{Warped Black String}) has an event horizon at $%
r_{+}=l$ with temperature $\beta =2\pi l$. This suggests that it could evaporate decaying by
Hawking radiation into states with negative mass. The possibility of decay, however, would be in
conflict with the fact that the massless state is supersymmetric.

A supersymmetric ground state\ with metric (\ref{Warped Black String}) is given by
\begin{equation}\label{The Wormhole}
ds^{2}= -\sinh ^{2}\left(\frac{w}{l}\right) dt^{2}+dw^{2}+ l^2 \cosh
^{2}\left(\frac{w}{l}\right)\frac{dz^{2}+d\varphi ^{2}}{z^{2}} ,
\end{equation}
with $-\infty <w<\infty$. This manifold $\mathcal{M}$ has negative constant curvature and is
smooth everywhere. It possesses a  horizon at $w=0$, where an Einstein-Rosen bridge is
centered,
and its boundary is formed by two connected components, defined by $%
w=\pm \infty $, so that $\partial \mathcal{M}$ is a manifold\ of negative scalar curvature. This
does not contradict the no-go theorem for wormholes in AdS\ of Ref. \cite{Witten:1999xp}, which\
states that a boundary with positive scalar curvature must be connected.

In this scenario an observer sees a \textit{warped black string }with $\mu
>0 $ decaying towards a supersymmetric final state (\ref{The Wormhole})
preserving half of the supersymmetries of AdS$_{4}$, generated by the Killing spinors

\begin{equation}
\epsilon =\exp \left(-\frac{w}{2l}\Gamma _{1}\right)\exp \left(\frac{t}{2l}\Gamma _{1}\Gamma
_{0}\right)\eta,
\end{equation}
with
\[
\eta =\left(
\begin{array}{l}
\exp (-\frac{\ln (z)}{2}\sigma _{2})\xi _{+} \\
\exp (\frac{\ln (z)}{2}\sigma _{2})\xi _{-}
\end{array}
\right) \;,
\]
where the constant two-dimensional spinors $\xi _{\pm }$ satisfy $\sigma _{2}\xi _{\pm }=\pm \xi
_{\pm }$. This ground state can be obtained from the exterior\ metric of (\ref{Warped Black
String}) with $\mu =0$, $r=l\cosh w/l$, and thereafter continuing $w$ to negative values.

Warped black branes of higher dimensions, analogous to (\ref{Warped Black String}), can be readily
found,
\begin{equation}\label{WarpeHigherBrane}
ds^{2} =-\left( r^{2}/l^{2}-\frac{\mu }{r^{d-3}}-1\right) dt^{2}+
\frac{dr^{2}}{\left(r^{2}/l^{2}-\frac{\mu }{r^{d-3}}-1\right) } +r^{2}d\sigma _{-1}\;,
\end{equation}
where the $d-2$ dimensional transverse sections $\Sigma _{-1}$ have metric
\begin{equation}
d\sigma _{-1}^{2}=\frac{1}{z^{2}}\left( dz^{2}+\delta _{ij}dx^{i}dx^{j}\right) \;,
\label{warpedHn}
\end{equation}
with $0<z<\infty $ and at least one of the $x^{i}$'s is compact. The mass is given by
$M=V_{d-2}\frac{\mu }{2\Omega _{d-2}}$ where $\Omega _{d-2}$ is the volume of $d-2$ sphere. The
ground state is given by the higher dimensional version of metric (\ref{The Wormhole}) with
transverse section (\ref {warpedHn}). In this case, the ground state preserves one half of the
supersymmetries of AdS$_{d}$, and the Killing spinors of $\Sigma _{-1}$ are
\begin{equation}
\eta _{\pm }=\exp \left(\mp \frac{\ln (z)}{2}\gamma _{2}\right)\xi _{\pm },
\end{equation}
where $(I\pm \gamma _{2})\xi _{\pm }=0$. This last condition comes from the\ periodicity in one of
the $x^{i}$'s. Note that these Killing spinors depend only on $z$, and therefore the coordinates
$x^{i}$ can be further wrapped without breaking additional supersymmetries \cite{Explaining}.

\subsection{Ricci flat transverse section}

\label{ZeroCurvature}

For $\gamma =0$, the solution of the Killing spinor equation (\ref{Solution}%
) reads
\begin{equation}
\epsilon =\exp \left( -\frac{\Gamma _{1}}{2}\ln \left(\frac{r}{l}\right) \right) \left(
1-\frac{t}{l}P_{0}\right) \eta \;,
\end{equation}
where $\eta $\ satisfies $\left( d+\hat{A}\right) \eta =0$. Here $\hat{A}=%
\frac{1}{2}\hat{\omega}^{mn}J_{mn}+\hat{e}^{m}P_{m}$, where $J_{mn}$ and $%
P_{m}$ generate $ISO(d-2)$, and $P_{0}$, $P_{m}$ can be read from Eq.(\ref {Ps}).\ Due to the
presence of the local translation generators $P_{m}$, the spinor $\eta $ does not necessarily
satisfy the standard Killing spinor equation,
\begin{equation}
d\eta +\frac{1}{4}\hat{\omega}^{mn}\gamma _{mn}\eta =0\;.
\end{equation}
The consistency conditions, however, still require the transverse section $%
\Sigma _{0}$ to be Ricci flat. Moreover, the integration constant $\mu $
could be rescaled away in the metric (\ref{Top-BH-Einstein}) with $\gamma =0$%
, unless one of the coordinates of the transverse section $y^{i}$ is compact. In order to avoid
this ambiguity, transverse geometries with one wrapped direction, of the form $\Sigma
_{0}=S^{1}\times \Xi $, shall be considered. The line element of $\Sigma _{0}$ reads,
\begin{equation}
d\sigma _{0}^{2}=d\phi ^{2}+h_{ij}(x)dx^{i}dx^{j}\;,  \label{FlatTransverse}
\end{equation}
where $h_{ij}$ is the metric of the $(d-3)$-dimensional Ricci flat Euclidean space $\Xi $, and
$\phi $ parametrizes $S^{1}$. In this case, the solution of the Killing equation is given by

\[
\eta =(1-\frac{1}{2}\phi \Gamma _{2}P_{+})\tilde{\eta}(x),
\]
where $\tilde{\eta}(x)$ satisfies the standard Killing equation on $\Xi $. The periodicity
condition implies that $P_{+}\tilde{\eta}=0$, which projects out half of the components of $\eta
$, and thus, the spinor $\eta _{-}=P_{-}\eta $ satisfies

\begin{equation}
d\eta _{-}+\frac{1}{4}\hat{\omega}^{mn}\gamma _{mn}\eta _{-}=0, \label{FlatEquation}
\end{equation}
where the representation $\Gamma _{0}=i\sigma _{y}\otimes \mathbb{I}$, $%
\Gamma _{1}=\sigma _{z}\otimes \mathbb{I}$, $\Gamma _{m}=\mathbb{\sigma }%
_{x}\otimes \gamma _{m}$ has been used. In this way, Killing spinors for the spacetime
(\ref{Top-BH-vacuum}) with $\gamma =0$ possess only one chirality. Now, since $\eta =\eta _{-}$
solves Eq. (\ref{FlatEquation}), the supersymmetric Ricci flat surfaces $\Xi $ are classified
\cite{Wang1989},\ therefore, black objects given by (\ref{Top-BH-Einstein}) with $\gamma =0$,
whose transverse sections have metric (\ref{FlatTransverse}), possessing a supersymmetric ground
state, can be classified according to the following theorem:

\textbf{Theorem 3.} Let $\mathcal{M}$ be a $d$-dimensional manifold of the form (\ref
{Top-BH-Einstein}) with $\gamma =0$, with transverse section of the form $%
\Sigma _{0}=S^{1}\times \Xi $, where $\Xi $ is a simply connected, complete and irreducible Ricci
flat manifold. If $\mathcal{M}$ possesses a globally supersymmetric ground state, then it can be
either

\textbf{(i)} a $d$-dimensional black brane with $\Sigma _{0}=S^{1}\times %
\mathbb{R}^{d-3}$, whose ground state ($\mu =0$), is a locally AdS spacetime admitting
$2^{[(d-3)/2]}$ Killing spinors,

\textbf{(ii)} a $10$-dimensional black object, where $\Xi $ is a $7$%
-dimensional space with $G_{2}$ holonomy admitting only one Killing spinor, or

\textbf{(iii)} an odd-dimensional black object with $d\geq 7$, where the geometry, holonomy and
corresponding maximal number of Killing spinors of $\Xi$ are the same as those listed in Table
$2$.

In higher dimensions, the number of Killing spinors, as well as the geometry of $\Xi $ can be
readily obtained from those in the\ previous section. Note that if\ $\Xi $\ is assumed to be non
simply connected, then in the maximally supersymmetric case, where $\Sigma _{0}=S^{1}\times \mathbb{R}%
^{2n-3}$, the remaining coordinates can be further wrapped, so that $\Sigma
_{0}=(S^{1})^{p+1}\times \mathbb{R}^{2n-3-p}$ with $0\leq p\leq 2n-3$, without breaking additional
supersymmetries.

\section{AdS Supergravity in eleven dimensions}

\label{Elevendimensions}

It has been shown that standard eleven dimensional supergravity \cite {Cremmer:1978km} cannot
accommodate a cosmological constant \cite {Bautier:1997yp}, so it would be interesting to examine
whether the supersymmetric solutions discussed here make sense in eleven dimensions. It turns out
that these solutions are BPS states of eleven dimensional AdS supergravity
\cite{Troncoso:1998va,Troncoso:1998ng}. The field content of the $\mathcal{N}=1$ theory is the
graviton $e_{\mu }^{a}$, a gravitino $\psi
_{\mu }$ , the spin connection $\omega _{\mu }^{ab}$, and a bosonic 1-form $%
b_{\mu }^{abcde}$ which is an antisymmetric fifth-rank Lorentz tensor in tangent space. These
fields form a connection for the super AdS$_{11}$ group, $OSp(32|1)$, whose algebra is expected to
be the underlying M-Theory symmetry. The action describes a gauge theory with a fiber bundle
structure, and the Lagrangian is a Chern-Simons density. The purely gravitational sector of the
Lagrangian contains a negative cosmological constant and the Einstein-Hilbert term,\ plus some
additional terms with higher powers of the Riemann curvature, combined in such a way as to yield
second order field equations for the metric.

This theory possesses solutions of the form \cite{Aros:2000ij}
\begin{equation}\label{Topo11}
ds^{2} =-\left( \gamma +r^{2}/l^{2}-(2G\mu )^{1/5}\right) dt^{2}+ \frac{dr^{2}}{\left( \gamma
+r^{2}/l^{2}-(2G\mu )^{1/5}\right) } +r^{2}d\sigma _{\gamma }^{2},
\end{equation}
where the integration constant $\mu $ is related to the mass through $\mu =%
\frac{\Omega _{9}}{V_{9}}M+\frac{1}{2G}\delta _{1,\gamma }$. These
configurations are left invariant under the supersymmetry transformation $%
\delta \psi =\nabla \epsilon $, provided $\epsilon $ solve the same the Killing spinor equation as
Eq.(\ref{KillingAdS} ) and, as a consequence, requiring supersymmetry restricts the transverse
section to be an Einstein manifold of scalar curvature $\hat{R}=72\gamma $. With this last
condition, the metric (\ref{Topo11}) solves the field equations, even though they are not those of
Einstein's theory.

The supersymmetric ground states correspond to the metric (\ref{Topo11}) with $\mu =0$, which for
$d=11$ is the same as Eq.(\ref{Top-BH-vacuum}), and therefore, the classification in eleven
dimensions can be obtained from the theorems above. If $\gamma =1$, the transverse section can be
either $S^{9}$ or a Sasaki-Einstein manifold. If $\gamma =-1,$ the transverse section is a
negative scalar curvature manifold of the form (\ref{WarpedEta}) with a subsection $\Xi $ which
can be either $\mathbb{R}^{8}$, hyperk\"{a}hler, Calabi-Yau, or a Parallel Spin$_{7}$ manifold.
Finally, if $\gamma =0$ and
the transverse section is $\Sigma _{0}=S^{1}\times \Xi $, then the surface $%
\Xi $ coincides with the previous case.

\section{Discussion}

Black objects of the form (\ref{Top-BH-Einstein}) possessing a supersymmetric ground state have
been classified. These geometries necessarily have a constant curvature transverse section $\Sigma
$, if the spacetime dimension is less than seven. For $d\geq 7$, the transverse section $\Sigma $
can be also any of the Euclidean Einstein manifolds listed in Tables $1$ and $2$. Since the
existence of manifolds with exceptional holonomy, even dimensional spacetimes with a nonconstant
curvature transverse section exist only for $d=8$, being $\Sigma $ a surface of positive scalar
curvature, and for $d=10$, with $\Sigma $ a nonpositive scalar curvature manifold.

In odd dimensions, this classification goes beyond standard supergravity, in particular, the
eleven dimensional case was analyzed in Section \ref {Elevendimensions}.

As it occurs for conifold geometries one would expect that besides the mapping of Killing spinors
between a supersymmetric ground state and its transverse section, further structures can be
connected. Following this scheme, one would expect that other BPS states as branes or product
spaces can be classified.

The spacetimes discussed here are asymptotically locally AdS, only when the transverse section has
constant curvature, such as the massless configuration (\ref{Top-BH-vacuum}) is a constant
curvature manifold. In this case, the curvature $F=dA+A\mbox{\tiny $\wedge$}A$ vanishes, so that
the AdS
connection $A$ in Eq.(\ref{A}) can be expressed in terms of an element of $%
SO(d-1,2)$ as $A=g^{-1}dg$. Hence, the Killing spinor equation is solved by
\begin{equation}
\epsilon =g^{-1}\epsilon _{0}\;,  \label{Epsilon}
\end{equation}
where $\epsilon _{0}$ is a constant spinor, and the group element reads
\begin{equation}
g(t,r,y^{m})=T(y^{m})e^{\frac{t}{l}P_{0}}e^{\frac{\Gamma _{1}}{2}\ln \left( r/l+\sqrt{\gamma
+r^{2}/l^{2}}\right) }\;,  \label{TBHg}
\end{equation}
with $T(y^{m})$ satisfying
\begin{equation}
dT(y^{m})=T(y^{m})\hat{A}_{\gamma }\;.  \label{TransverseEquation}
\end{equation}
Here $\hat{A}$ is a flat connection for the transverse section, which means that the Killing
spinors of $\Sigma $ are given by $\eta =T^{-1}\epsilon _{0} $. In the case $\gamma =-1$, the
temperature, $\beta =2\pi l$, can be found either by demanding regularity of the Euclidean
solution at the horizon or by demanding the holonomy of $A$ to be trivial, that is $g^{-1}(\beta
,r,y)g(0,r,y)=1$, where $g(\tau ,r,y)$ is obtained from Eq.(\ref{TBHg}) by a Wick rotation.
Killing spinors for other locally AdS spacetimes have been discussed also \cite
{Banados:1998dc,Ghosh:1999nf,Kehagias:2000dg,Shuster:1999zf}.

For $\gamma =-1$ the requirement of global supersymmetry implies that only extended objects can
have a supersymmetric ground state. An explicit example resembling a wormhole was constructed (see
Eqs.(\ref{Warped Black String}, \ref{The Wormhole},\ref{WarpeHigherBrane})), which is a
supersymmetric state with nonvanishing temperature as it occurs for some BPS branes in Ref. \cite
{Sen:1995in}.

If the transverse section were a Ricci flat manifold which differs from $%
\Sigma _{0}=S^{1}\times \Xi $, the classification deals with a different problem. In that case,
for $\gamma =0$, the equation (\ref{KillingTransverse} ) on the transverse section does not reduce
in general to the standard Killing spinor equation. In fact, using the representation $\Gamma
_{0}=i\sigma _{y}\otimes \mathbb{I}$, $\Gamma _{1}=\sigma _{z}\otimes %
\mathbb{I}$, $\Gamma _{m}=\mathbb{\sigma}_{x}\otimes \gamma _{m}$, the spinors $\eta _{\pm
}=P_{\pm }\eta $\ satisfy

\begin{equation}
d\eta _{+}+\frac{1}{4}\hat{\omega}^{mn}\gamma _{mn}\eta _{+}=0\;, \label{FlatEquation1}
\end{equation}
and
\begin{equation}
d\eta _{-}+\frac{1}{4}\hat{\omega}^{mn}\gamma _{mn}\eta _{-}=\frac{1}{2}%
\gamma _{m}\hat{e}^{m}\eta _{+}\;.  \label{FlatEquation2}
\end{equation}
The solutions for $\eta _{-}$ can be trivially written in terms of $\eta _{+} $. Indeed, the
consistency condition for Eq.(\ref{FlatEquation2}) gives the same information as
Eq.(\ref{FlatEquation1}), \emph{i. e.}, $\Sigma _{0}$ must be Ricci flat. However, equation
(\ref{FlatEquation2}) may be incompatible with some of the global properties of $\Sigma _{0}$, and
hence\ it would in general restrict the possible spaces where Eq.(\ref {FlatEquation1}) has a
global solution.

\acknowledgments

We would like to thank Professors  Christian B\"{a}r, Helga Baum, Gary Gibbons, Marc Henneaux and
Claudio Teitelboim for helpful comments. This research is partially funded by FONDECYT grants
3990009, 1010446, 1010449, 1010450, 1020629, 7010446, 7010450 and for the generous support to CECS
by Empresas CMPC. CECS is a Millennium Science Institute.

\providecommand{\href}[2]{#2}\begingroup\raggedright\endgroup


\begin{thebibliography}{99}
\bibitem{Explaining}  The Killing spinors of $H^{d-2},$ written in terms of
Poincar\'{e} coordinates, are
\[
\eta =\left( e^{-\frac{1}{2}\ln (z)\gamma _{2}}\prod_{i=3}^{d-1}\left( 1-%
\frac{1}{2}x^{i}\gamma _{i}(I+\gamma _{2})\right) \right) \eta _{0}.
\]
For the transverse section (\ref{warpedHn}) the compactness of a single $%
x^{i}$ implies the projection condition $(I+\gamma _{2})\eta _{0}=0$, and additional wrappings
along the other directions do not impose further constraints.
\end{thebibliography}
\end{document}